\documentclass[aps,prl,twocolumn,superscriptaddress,amsmath,amssymb,nofootinbib]{revtex4-2}
\usepackage{graphicx}
\usepackage{babel}
\usepackage{color}
\usepackage{bbm}
\usepackage{hyperref}
\usepackage{comment}
\hypersetup{colorlinks,urlcolor=blue}

\begin{document}

    \title{Exchange gap in GdPtBi probed by magneto-optics}

    \author{S.~Polatkan}
    \affiliation{1.~Physikalisches Institut, Universit\"at Stuttgart, Pfaffenwaldring 57, 70569 Stuttgart, Germany}
    \author{E.~Uykur}
    \affiliation{1.~Physikalisches Institut, Universit\"at Stuttgart, Pfaffenwaldring 57, 70569 Stuttgart, Germany}
    \affiliation{Helmholtz-Zentrum Dresden-Rossendorf, Institute of Ion Beam Physics and Materials Research, 01328 Dresden, Germany}
    \author{I.~Mohelsky}
    \affiliation{Laboratoire National des Champs Magn\'etiques Intenses (LNCMI), CNRS-UGA-UPS-INSA-EMFL, 38042 Grenoble, France}
    \author{J.~Wyzula}
    \affiliation{Laboratoire National des Champs Magn\'etiques Intenses (LNCMI), CNRS-UGA-UPS-INSA-EMFL, 38042 Grenoble, France}
    \affiliation{Department of Physics, University of Fribourg, 1700 Fribourg, Switzerland}
    \author{M.~Orlita}
    \affiliation{Laboratoire National des Champs Magn\'etiques Intenses (LNCMI), CNRS-UGA-UPS-INSA-EMFL, 38042 Grenoble, France}
    \affiliation{Faculty of Mathematics and Physics, Charles University, Ke Karlovu 5, Prague, 121 16, Czech Republic}
    \author{C.~Shekhar}
    \author{C.~Felser}
    \affiliation{Max-Planck-Institut f\"{u}r Chemische Physik fester Stoffe, 01187 Dresden, Germany}
    \author{M.~Dressel}
    \author{A.~V.~Pronin}
    \affiliation{1.~Physikalisches Institut, Universit\"at Stuttgart, Pfaffenwaldring 57, 70569 Stuttgart, Germany}

    \date{\today}

	\begin{abstract}
		We measured the magneto-reflectivity spectra (4 -- 90 meV, 0 -- 16 T) of the triple-point semimetal GdPtBi and found them to demonstrate two unusual broad features emerging in field. The electronic bands of GdPtBi are expected to experience large exchange-mediated shifts, which lends itself to a description via effective Zeeman splittings with a large $g$ factor. Based on this approach, along with an \textit{ab initio} band structure analysis, we propose a model Hamiltonian that describes our observations well and allows us to estimate the effective $g$ factor, $g^{*} = 95$. We conclude that we directly observe the exchange-induced $\Gamma_{8}$ band inversion in GdPtBi by means of infrared spectroscopy.
	\end{abstract}

\maketitle

\begin{figure}[]
	\includegraphics[width=1\linewidth]{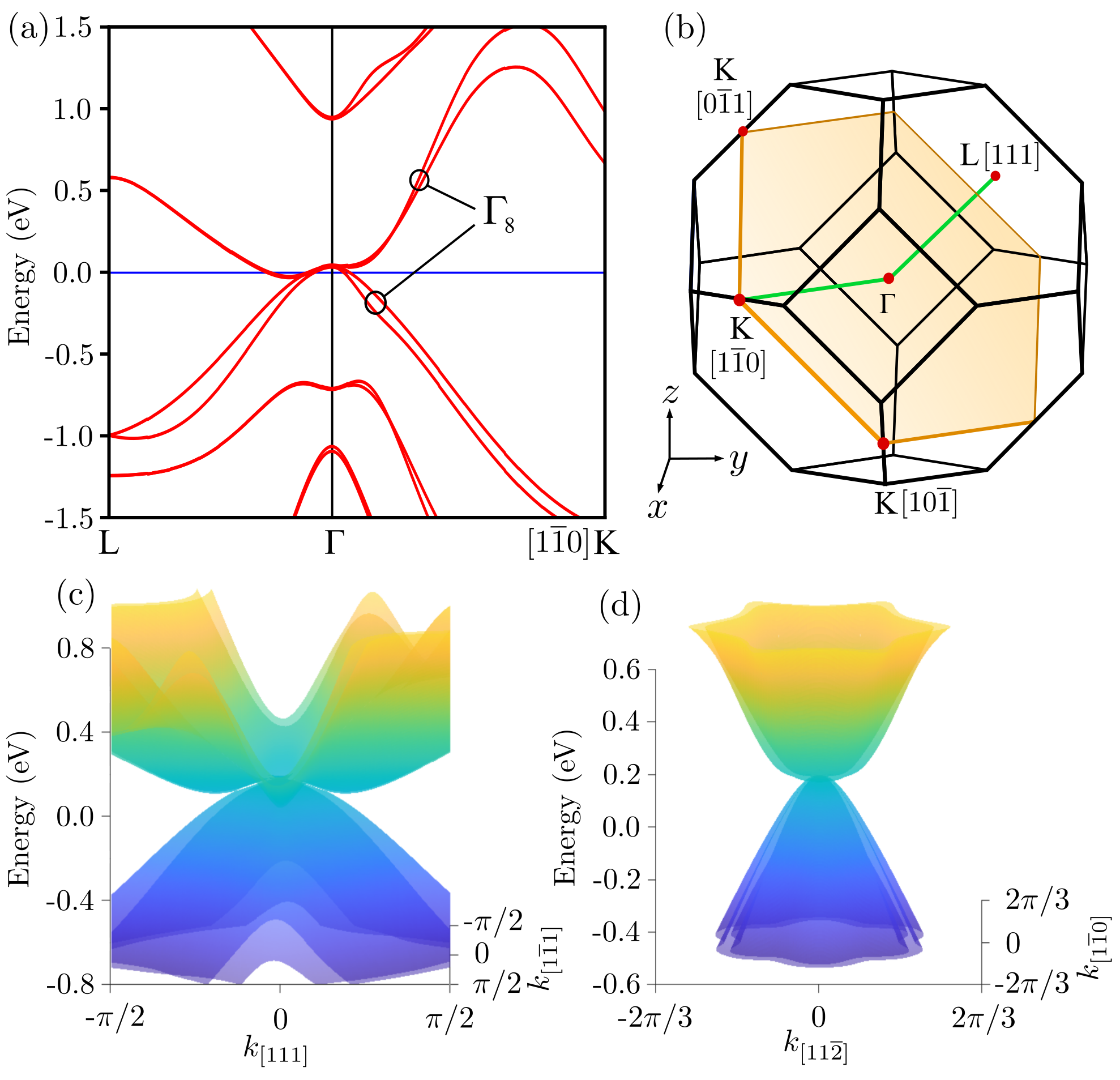}
	\caption{(a) Band structure of GdPtBi along a $L-\Gamma-K$ line and (b) the compound's Brillouin zone (BZ) showing this line path and a (111) plane, relevant to the measurements (b). 2D cuts of the BZ, containing (c) the [111] line and (d) within the (111) plane. Additional band-structure plots are provided in the Supplemental Material~\cite{SM}.}
	\label{FigConstrained}
\end{figure}

\textit{Introduction.}
Half-Heusler compounds are a family of ternary cubic crystals that provide a large variability in their band structures and hence in their material properties. These materials can host inverted band gaps, leading to topologically non-trivial electronic states~\cite{Chadov2010, Lin2010}. This paper focuses on GdPtBi -- a magnetic half-Heusler material with a semimetallic electronic structure and an antiferromagnetic phase below the N\'eel temperature, $T_\textrm{N}=9$~K \cite{Canfield1991}. In this and related compounds, the $s$-$p$ band inversion leads to degenerate states at the $\Gamma$ point, creating a pair of a W-shaped electron bands resting on a parabolic hole band~\cite{Suzuki2016, Yang2017}. Upon close inspection, these touching bands cross each other, forming the so-called triple points, a type of topological intersection where a doubly degenerate band intersects a pair of split bands at a point in $k$ space \cite{Winkler2019}. Such an intersection is realized in GdPtBi on the $L - \Gamma - L$ symmetry line; see Fig.~\ref{FigConstrained}(a). This band structure exists in a variety of half-Heusler compounds \cite{Chadov2010, Yang2017} and has been experimentally supported by angle-resolved photoemission spectroscopy (ARPES) and other methods~\cite{Liu2016c, Guo2018, Shekhar2018, Hutt2018, Schindler2020}.

When GdPtBi is subjected to a magnetic field, the $\Gamma_8$ bands are believed to substantially split in a Zeeman-like fashion~\cite{Hirschberger2016, Cano2017}, with large effective $g$ factors mediated by the exchange interaction~\cite{Hirschberger2016, Cano2017, Shekhar2018}. The splitting of the triple points in fields gives rise to Weyl points, which positions depend on the field strength.

This magnetically tunable Weyl state in GdPtBi has attracted quite a lot of attention. Due to the inaccessibility of ARPES in magnetic fields, other experimental means are required to investigate the exchange-split band structure of GdPtBi. Currently, electrical and thermal magnetotransport measurements are available~\cite{Shekhar2018, Hirschberger2016, Schindler2020}. Magneto-optical studies offer another route for such investigations, but are still missing. In this paper, we amend this shortage. Our findings provide strong and independent experimental evidence for the large effective Zeeman splitting in GdPtBi.\\

\textit{Methods.} GdPtBi single crystals were grown by the solution method from a Bi flux. Freshly polished pieces of Gd, Pt, and Bi, each of purity larger than 99.99\%, in the ratio Gd:Pt:Bi = 1:1:9 were placed in a tantalum crucible and sealed in a dry quartz ampoule under 3 mbar partial pressure of argon. The filled ampoule was heated at a rate of $100~\text{K/hr}$ up to $1200^\circ$C, followed by 12 hours of soaking at this temperature. For crystal growth, the temperature was slowly reduced by $2~\text{K/hr}$ to $600^\circ$C. Extra Bi flux was removed by decanting it from the ampoule at $600^\circ$C. Overall, the crystal-growth procedure closely followed the one described in Refs.~\cite{Shekhar2018, Canfield1991}. The crystals’ composition and structure (noncentrosymmetric F$\overline{4}$3m space group) were checked by energy dispersive $x$-ray analysis and Laue diffraction, respectively.

The magneto-optical data were collected in reflectivity mode from a (111)-oriented facet with the area of roughly 2 by 1.1 mm$^{2}$ (same sample as used for the broadband optical measurements without a magnetic field~\cite{Hutt2018}). The sample was kept at $T=4.2$~K, which is below $T_\textrm{N}$, in helium exchange gas during the measurements. The sample holder was placed in a superconducting coil, which provided magnetic fields up to 16 T. We employed the Faraday configuration, i.e., the field was parallel to the light propagation direction and to the [111] crystallographic direction; see, also, Fig.~\ref{FigConstrained}(b) for the corresponding cut of the Brillouin zone (BZ). Far-infrared radiation ($\sim4 - 90$ meV) from a Hg lamp (below $45~\text{meV}$) or a globar (above $45~\text{meV}$) was delivered to the sample via light-pipe optics. The reflected light was directed to a Bruker Vertex80v Fourier-transform spectrometer and detected by a liquid-helium-cooled bolometer placed outside the magnet. The sample’s reflectivity $R_{\rm{B}}$ at a given magnetic field $B$ was normalized by the sample’s reflectivity $R_0$ measured at $B = 0$.

The density-functional theory (DFT) calculations were performed with the WIEN2K code \cite{Blaha2020}, which is based on the (linearized) augmented plane-wave and local orbitals [(L)APW+lo] method to solve the Kohn-Sham equations \cite{Kohn1965} of DFT. The exchange-correlation potential was calculated using the Perdew-Burke-Ernzerhof generalized gradient approximation~\cite{Perdew1996}. Results were crosschecked with runs performed on a $40 \times 40 \times 40$ $k$ mesh, with the energy converged below $10^{-8}$~Ry and charge converged to $10^{-8}~e$ for the calculations without spin-polarization and a $50 \times 50 \times 50$ $k$ mesh converged below $10^{-7}$~Ry and $10^{-7}~e$.

\begin{figure*}
	\includegraphics[width=\linewidth]{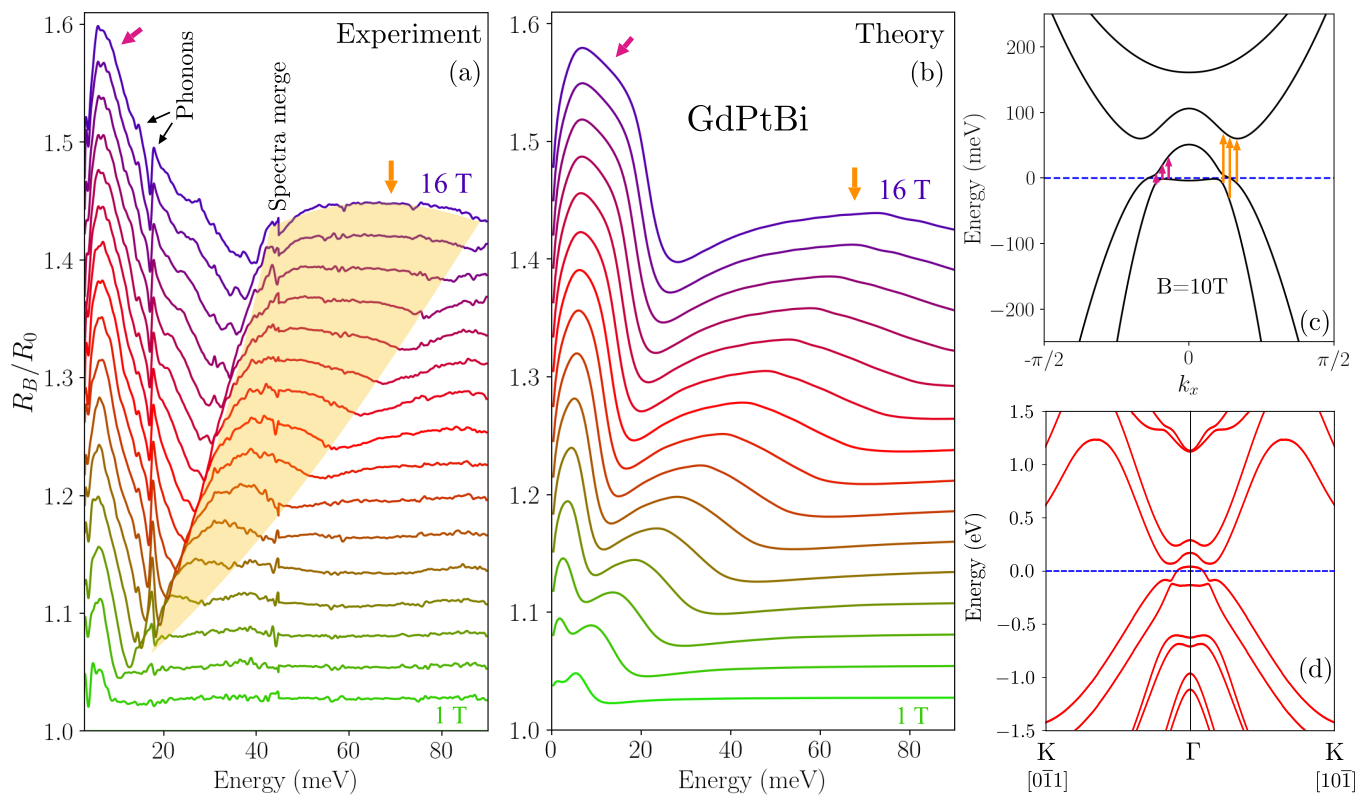}
	\caption{(a) Relative magneto-reflectivity spectra of GdPtBi measured from the (111) plane at $T=4.2$~K and various magnetic fields from 1 to 16 T. (b) Spectra of the same type, calculated from the four-band model Hamiltonian, given by Eq.~\eqref{HamiltonianEq}. (c) The bands of the four-band Hamiltonian for $B=10~\text{T}$. (d) DFT band structure within the (111) plane at full spin polarization. Note that the field-induced Weyl points are not within this plane. The colored arrows indicate the major spectral features in (a) and (b) and the corresponding interband transitions in (c). The transitions between the two hole bands near the gapped nodal line produce the low-energy peak, while the transitions to the electron bands are responsible for the high-energy maximum; see the Supplemental Material~\cite{SM} for details.}
	\label{FigResults}
\end{figure*}

\textit{Results and discussion.} The experimental results are presented in Fig.~\ref{FigResults}(a) as a stacked plot of $R_{\rm{B}}/R_0$ in 1~T steps. The spectra are dominated by two broad peaks, emerging with magnetic field: a low-frequency peak below 10~meV and a broad, less intense maximum, which shifts to higher frequencies as $B$ increases. This feature is marked by the transparent orange overlay in the figure. Other smaller visible features include two phonons at around 14.5 and 18~meV (in agreement with the phonons reported in Ref.~\cite{Hutt2018}) and an artifact at 45~meV, which appears due to the merger of two spectral ranges in our measurements. In this paper, we focus on the two broad peaks.

By looking at the raw spectra, one can immediately note that they are uncharacteristic for inter-Landau-level transitions, which typically manifest themselves as a rich series of rather narrow absorption lines in $R_{\rm{B}}/R_0$ (``Landau fans''), as reported in numerous studies of different (topological) semimetals performed by our and other groups~\cite{ChenZ2017, Jiang2018, Yuan2018, Polatkan2020, Wyzula2022, Santos2022, Mohelsky2023}. The inter-Landau-level transitions in GdPtBi are likely weak and not resolved in our measurements.

The cyclotron resonance (CR) due to free carriers might manifest itself in $R_{\rm{B}}/R_0$, but is not expected to provide a dominating contribution to the spectra of our GdPtBi sample, as it has low carrier concentration, low plasma frequency, and low electron scattering rate at $T < 100$~K \cite{Hutt2018}. Indeed, as we demonstrate in the Supplemental Material~\cite{SM}, the CR provides a contribution at low frequencies, but the corresponding spectral feature is rather narrow and cannot describe (even qualitatively) the entirety of the experimental data. Thus, interband transitions must be considered as the major reason for the observed broad peaks.

As pointed out previously, the antiferromagnetic ordering of Gd spins in GdPtBi is soft and smoothly transitions into a collinear alignment when a magnetic field is applied~\cite{Shekhar2018}. The band structure strongly changes with this alignment: upon application of a magnetic field, the spin-nondegenerate bands near the $\Gamma$ point start to diverge in energy in a Zeeman-like fashion due to the exchange interaction~\cite{Suzuki2016, Hirschberger2016, Cano2017, Yang2017}. In the following, we demonstrate that the two broad features in our $R_{\rm{B}}/R_0$ spectra can be well modeled by the transitions within the exchange-interaction split bands.

To illustrate how the exchange-shifted bands can give rise to the observed features, we first discuss a model Hamiltonian. We adopt the notion of effective Zeeman terms, proportional to spin-$\frac{3}{2}$ matrices, employed in the $k \cdot p$ model of Ref.~\cite{Cano2017}. However, we find that this model does not reproduce the DFT band structure well in the (111) plane (which we probe in our measurements) and propose a toy model instead, which bears the characteristic features of the bands. Thus, our model Hamiltonian reads:
\onecolumngrid
\begin{equation}
	H= \begin{pmatrix}
		\frac{\hbar^2 \mathbf{q}^2}{2 m_{e}} +\mu_{\frac{3}{2}} B& 	\frac{1}{2}Y_- & 		\frac{1}{2}Y_+ & 		0 \\
		\frac{1}{2}Y_- & 		\frac{\hbar^2 \mathbf{q}^2}{2 m_{h}} + \mu_{\frac{1}{2}} B& 		Y_- & 	-\frac{1}{2}Y_+ \\
		\frac{1}{2}Y_+ & 		Y_- & 	\frac{\hbar^2 \mathbf{q}^2}{2 m_{h}} +\mu_{-\frac{1}{2}} B& 		\frac{1}{2}Y_- \\
		0 & 		-\frac{1}{2}Y_+ & 		\frac{1}{2}Y_- & 	\frac{\hbar^2 \mathbf{q}^2}{2 m_{e}} + \mu_{-\frac{3}{2}} B
	\end{pmatrix} + \Delta \cdot B, ~~~ Y_{\pm}=-\frac{\hbar^2}{2 m_{xy}} \left(q_x^2 \pm q_y^2 \right).
	\label{HamiltonianEq}
\end{equation}
\twocolumngrid

\begin{figure}[b]
	\includegraphics[width=0.95\linewidth]{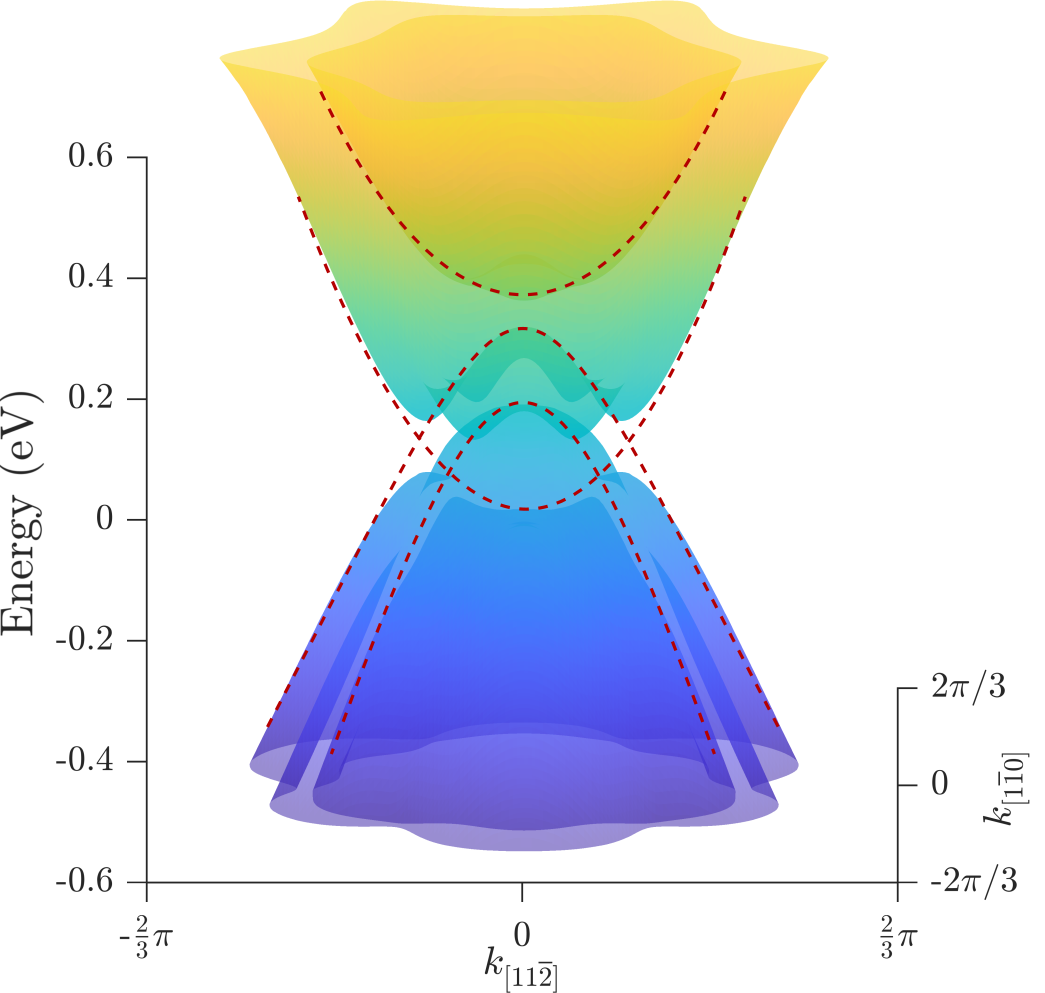}
	\caption{Two-dimensional DFT band structure of the (111) plane with the spins fully aligned along [111]. Exchange interaction splits the bands, which subsequently hybridize. Within the (111) plane, the electron and hole bands gap out, while between the two hole bands, a pocket forms. Dashed red lines schematically show the bands split by the exchange interaction, but not hybridized. The effective Zeeman terms in our toy model, given by Eq.~\eqref{HamiltonianEq}, correspond to these splittings (electron bands: $m_{s}=\pm\frac{3}{2}$; hole bands: $m_{s}=\pm\frac{1}{2}$). Additional renderings of the band structure are provided in the Supplemental Material~\cite{SM}.}
	\label{FigUnconstrained}
\end{figure}

It comprises two pairs of parabolic bands on the diagonal, describing electrons and holes with effective masses $m_e$ and $m_h$. As usual, $\mathbf{q}=(q_{x}, q_{y}, q_{z})$, $q_{i} = k_{i}/a$ with $a$ the lattice constant (recall that GdPtBi has a cubic symmetry), $i=\{x,y,z\}$, and $k_{i} \in [-\pi, \pi)$. The terms $\mu_{m_s} B$ provide the Zeeman-like splitting of the degenerate bands ($\mu_{m_s}= m_s g^* \mu_B$, where $m_s=\pm \frac{3}{2}, \pm \frac{1}{2}$, $g^*$ is an effective $g$ factor, and $\mu_B$ is the Bohr magneton). Due to this splitting, one electron band intersects with both hole bands, forming nodal lines. The off-diagonal terms $Y_{\pm}$ provide coupling between the bands. The $Y_+$ term is radially symmetric and introduces a gap between the electron and hole bands. A new nodal line emerges between the two hole bands. The $Y_-$ term couples these hole bands with one another, splitting the new nodal line with the exception of four points. It also couples the electron and hole bands to form the protruding flanks, qualitatively emulating the protrusions in the DFT band structure of GdPtBi; see Fig.~\ref{FigUnconstrained}. We take $m_{xy} = m_{0}/2$. Finally, $\Delta$ shifts all bands in energy, allowing one to tune the Fermi-level position $E_{F}$, $\Delta = 1.4 g^* \mu_B$. We kept $E_{F}$ near the band intersections [cf. Figs.~\ref{FigResults}(c) and \ref{FigResults}(d)] because our magneto-optical spectra do not demonstrate any gaplike features, which might appear due to the Pauli blocking and can be related to $E_{F}$. This is also consistent with the earlier optical-conductivity measurements performed on the same sample~\cite{Hutt2018}.

A cut through the energy spectrum of Eq.~\eqref{HamiltonianEq} is provided in Fig.~\ref{FigResults}(c). As one can see from the comparison with Fig.~\ref{FigResults}(d), where the DFT calculations are shown, the model reproduces the essential features of the low-energy band structure of GdPtBi quite well. The DFT calculations have been made for the magnetic field applied along the [111] direction at full spin polarization. Note that two inequivalent types of $K$ points, related by $C_{3}$ symmetry, exist in the (111) plane (cf. Fig.~\ref{FigConstrained}). The Fermi level has not been adjusted in the provided DFT calculations. The zero-field data though suggest a Fermi-level shift down in energy for our sample~\cite{Hutt2018}, making the correspondence between our model and the DFT calculations even better. Based on the model, we can thus assign the two broad peaks to the transitions indicated by the two sets of arrows given in Fig.~\ref{FigResults}(c).

The relative reflectivity spectra calculated from Eq.~\eqref{HamiltonianEq} are shown in Fig.~\ref{FigResults}(b). The calculations are made in a common way, using the Kubo formula and the standard optical formulas for reflectivity. The free-electron CR contribution is also included in Fig.~\ref{FigResults}(b) -- it provides a narrow peak on top of the low-energy maximum. Details of these calculations can be found in the Supplemental Material~\cite{SM}. As one can see from the figure, the model of Eq.~\eqref{HamiltonianEq} can qualitatively reproduce the two major features of the experimental $R_{\rm{B}}/R_0$ spectra and their $B$-field evolution: the low-frequency peak grows in intensity with increasing $B$ and remains roughly at a fixed frequency; the second peak shifts to higher frequency and broadens with field, centered around 70 meV at 16 T. Furthermore, the heights of both peaks in the calculated spectra are well comparable to the experiment.

Strictly speaking, the Hamiltonian should contain linear terms, or even cubic terms, to introduce inversion asymmetry. These are not included to keep the number of parameters limited. Further, including the linear terms would strongly skew the bands. Cano~\textit{et al.}~\cite{Cano2017} estimate the linear coefficient to be two orders of magnitude smaller than the parabolic ones in their $k\cdot p$ Hamiltonian, in line with our approach.

In the model, we used $m_{e} = \frac{3}{2} m_0,~ m_{h}= -\frac{1}{4} m_0$, which is in accord with previous measurements: estimates for the effective masses obtained from Shubnikov--de Haas oscillations range between $0.23~m_0$ \cite{Hirschberger2016} and $0.30~m_0$ \cite{Schindler2020}, while Hall measurements performed on slightly electron-doped GdPtBi in low fields yield an effective mass of $1.8~m_0$ \cite{Hirschberger2016}.

The values of the magnetic moments, $\mu_{\pm \frac{1}{2}, \frac{3}{2}}$, obtained within our model, lead to a large effective $g$ factor, $g^*=95$. As pointed out in the Introduction, such large values of $g^*$ have been anticipated for GdPtBi~\cite{Hirschberger2016, Cano2017}, but not detected experimentally. Note that $g^*$ expresses, in simple terms, the energy shift of the exchange-split bands as $B$ increases. The exchange-mediated effective $g$ factors have been discussed, e.g., in narrow-gap and inverted-gap semiconductors doped with magnetic ions \cite{Gaj1978, Bastard1981, Hoerstel1999}. From the DFT calculations, one can also extract an estimate for the $g$ factor. The energy gap between the $m_s=\pm 3/2$ bands at full spin alignment, which is experimentally reached at $B_{\upuparrows} = 25$ T \cite{Shekhar2018}, is roughly $\Delta E = 400$~meV (cf. Fig.~\ref{FigUnconstrained}), yielding $g^* = \Delta E/2\mu_B m_s B_{\upuparrows} \approx 92$, in good agreement with our model.

\textit{Conclusions.} We experimentally studied the magneto-optical reflectivity of the triple-point semimetal GdPtBi and found two broad maxima emerging in an applied magnetic field. We proposed a four-band model Hamiltonian, which mimics the low-energy DFT band structure and allows field-dependent calculations of the magneto-optical response. The model describes the field evolution of the bands near the Fermi level: the bands, degenerate at the $\Gamma$ point, split in a Zeeman-like fashion in applied fields due to the exchange interaction. The resulting gaps increase as the Gd $4f^7$ electrons align paramagnetically with field. The match between the experimental and computed magneto-optical spectra evidence the validity of the exchange mechanism proposed in Refs.~\cite{Shekhar2018, Hirschberger2016, Cano2017} for the magnetic-field evolution of the electronic bands in GdPtBi.

\textit{Acknowledgments.}
We thank Gabriele Untereiner for technical assistance and Mark O. Goerbig and Alexander Yaresko for fruitful discussions. This research was funded in part by the DFG via Grant No. DR228/51-3 and by a joint German-French Program for Project-Related Personal Exchange by DAAD and ANR via Project No. 57512163.

\bibliographystyle{apsrev4-2}
\bibliography{references}

\newpage

\newpage
\vspace*{-2.0cm}
\hspace*{-2.5cm} {
  \centering
  \includegraphics[width=1.2\textwidth, page=1]{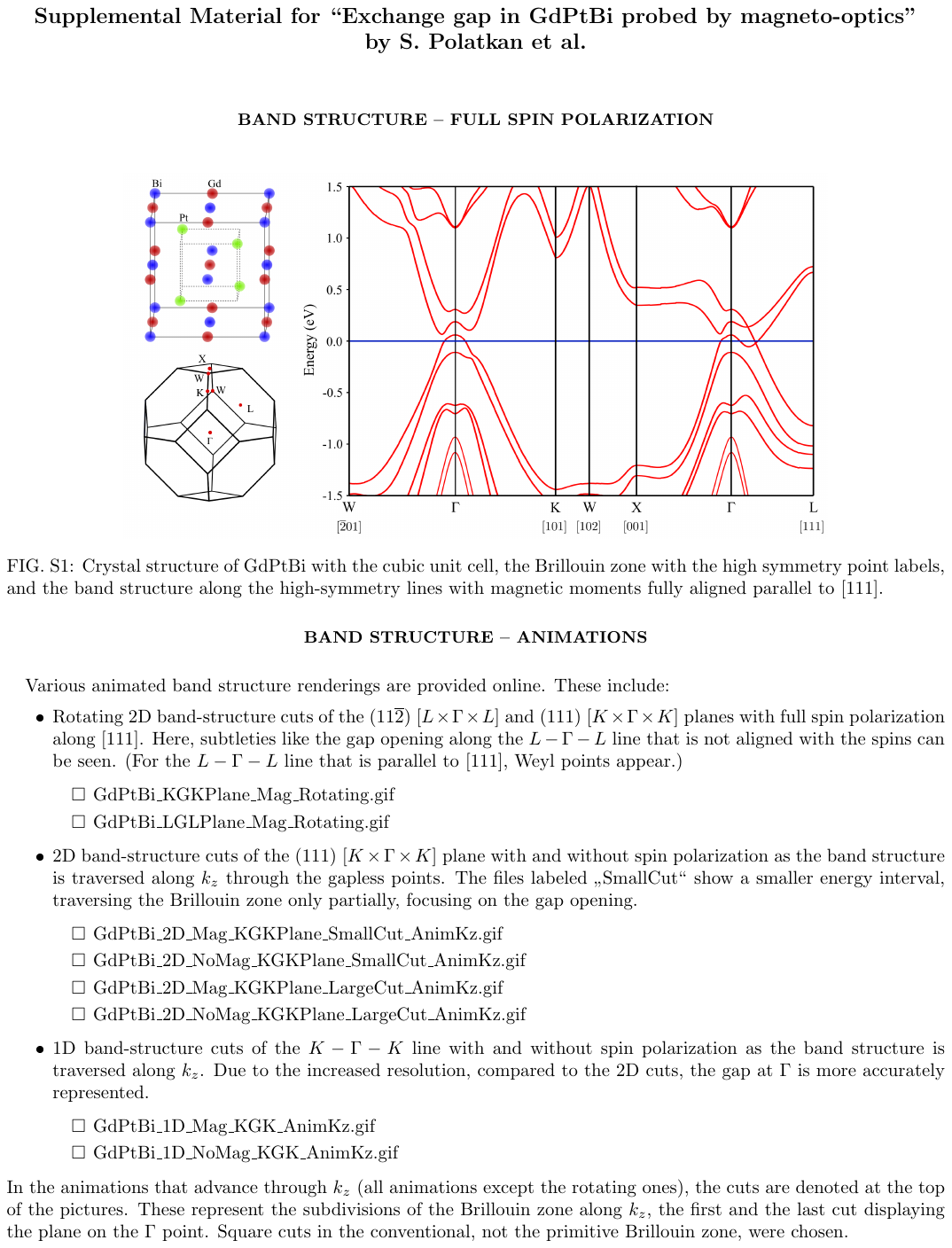} \\
  \ \\
}

\newpage
\vspace*{-2.0cm}
\hspace*{-2.5cm} {
  \centering
  \includegraphics[width=1.2\textwidth, page=2]{sm.pdf} \\
  \ \\
}

\newpage
\vspace*{-2.0cm}
\hspace*{-2.5cm} {
  \centering
  \includegraphics[width=1.2\textwidth, page=3]{sm.pdf} \\
  \ \\
}

\newpage
\vspace*{-2.0cm}
\hspace*{-2.5cm} {
  \centering
  \includegraphics[width=1.2\textwidth, page=4]{sm.pdf} \\
  \ \\
}

\end{document}